\documentclass[12pt,prb,reprint,aps,nofootinbib]{revtex4-1}
\usepackage{amssymb} \usepackage{amsmath} \usepackage{url}
\usepackage{graphicx} 
\usepackage[usenames, dvipsnames]{color}
\usepackage{cprotect}
\usepackage[normalem]{ulem}
\usepackage{float}
\newcommand{\N}{\mathbb{N}} \newcommand{\K}{\mathbb{K}}
\newcommand{\D}{\mathbb{D}} \newcommand{\A}{\mathbb{A}}
 \newcommand{\Z}{\mathbb{Z}}

\newcommand*{\algorithm}{\fontfamily{pcr}\selectfont}

\newcommand{\kb}{$\mathbf{k}$}

\graphicspath{{.}}

\begin{document}
\title{A robust algorithm for \kb-point grid generation and symmetry
  reduction} \author{Gus L. W. Hart} \affiliation{Dept. of Physics and
  Astronomy, Brigham Young University, Provo Utah 84602 USA}
\author{Rodney W. Forcade} \affiliation{Dept. of Mathematics, Brigham
  Young University, Provo Utah 84602 USA} \author{Jeremy J. Jorgensen}
\author{Wiley S. Morgan} \affiliation{Dept. of Physics and Astronomy,
  Brigham Young University, Provo Utah 84602 USA}

\begin{abstract}
We develop an algorithm for i) computing \emph{generalized regular}
\kb-point grids, ii) reducing the grids to their symmetrically distinct
points, and iii) mapping the reduced grid points into the Brillouin
zone. The algorithm exploits the connection between integer matrices
and finite groups to achieve a computational complexity that is linear
with the number of \kb-points. The favorable scaling means that, at a
given \kb-point density, all possible commensurate grids can be generated (as
suggested by Moreno and Soler) and quickly
reduced to identify the grid with the fewest symmetrically
unique \kb-points. These optimal grids provide significant speed-up
compared to Monkhorst-Pack \kb-point
grids; they have better symmetry reduction resulting in fewer irreducible \kb-points at a given grid density. The integer
nature of this new reduction algorithm also simplifies issues with
finite precision in current implementations. The algorithm is
available as open source software.
\end{abstract}
\maketitle
\section{Introduction}

\begin{figure}
\includegraphics[width=\linewidth]{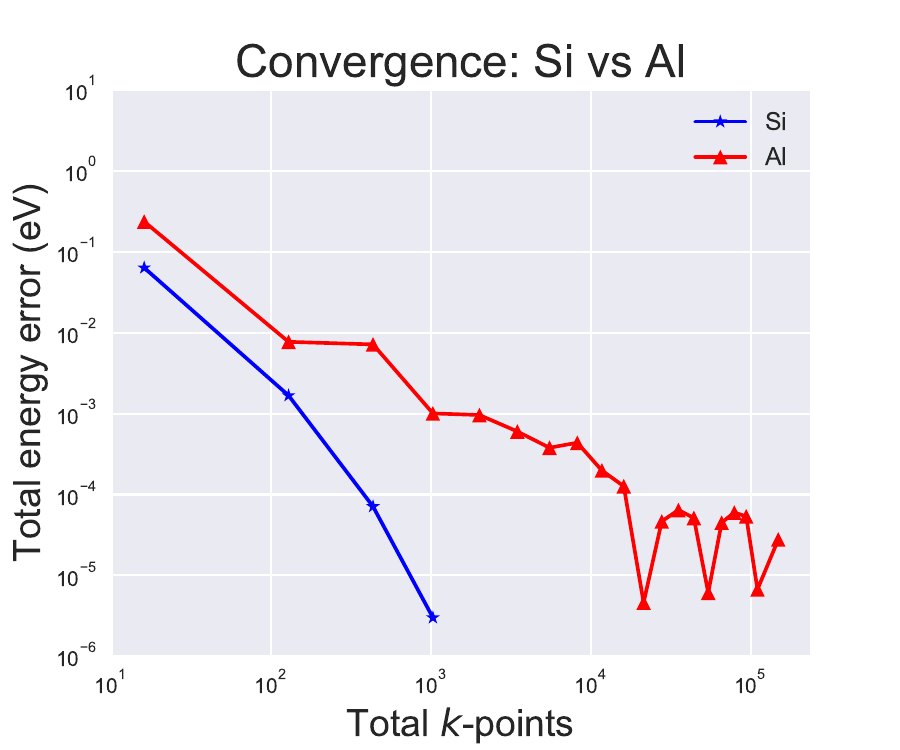}
\caption{Total energy error vs. \kb-point density for the cases of
  silicon and aluminum. Silicon does not have a Fermi surface so there
  is no discontinuity in the occupied bands; convergence is
  super-exponential or $\mathcal{O}(e^n)$ where $n$ is the number of
  \kb-points. (See the discussion of example 1 in
  Ref.~\onlinecite{weideman2002numerical}.) In contrast, the total
  energy of aluminum converges very slowly, and the convergence is
  quite erratic. For typical target accuracies in the total energy,
  around $10^{-3}$ eV/atom, metals require 10--50 times more
  \kb-points than semiconductors.}
\label{fig:convergence}
\end{figure}

Codes that solve the many-body problem using density functional theory (DFT)
use uniform grids over the Brillouin zone in order to calculate the
total electronic energy, among other material properties. The total
electronic energy is calculated by numerically integrating the
occupied electronic bands. For metallic systems, there
exist surfaces of discontinuities at the boundary between occupied and
unoccupied states, collectively known as the Fermi surface. These
discontinuities cause the accuracy in the calculation of the total
electronic energy to converge extremely slowly and erratically with
respect to grid density. This is demonstrated in
Fig. \ref{fig:convergence} where we compare the convergence of an
insulator (silicon) with a metal (aluminum).

The poor convergence of the electronic energy means that DFT codes
must use extremely dense
grids\cite{wisesa2016efficient,morgan2018efficiency} to achieve an
accuracy of several meV/atom.
To reduce computation time, it is common practice to evaluate eigenvalues at symmetrically equivalent \kb-points only once. This is the essence of ``symmetry reducing'' a \kb-point grid.

In most DFT codes, even for very dense grids, the setup and symmetry
reduction of the grid takes a few seconds at most.  Our motivation for
an improved algorithm (despite the speed of current routines) is
two-fold: 1) enable an automatic
grid-generation technique that allows us to scan over thousands of
candidate grids, in a few seconds, to find one with the best possible
symmetry reduction\cite{moreno1992optimal,froyen1989brillouin} (in
other words, enable a \kb-point generation method in the same
spirit as that of Ref.~\onlinecite{wisesa2016efficient} but have
the grid generation done \emph{on-the-fly} \cite{morgan2019generalized}),
and 2) eliminate (or at least greatly reduce) the probability of
incorrect symmetry reduction\footnote{Such errors are not uncommon in
  \kb-point reduction, but are not documented in the literature. The
  same errors are known to affect symmetry analysis as discussed at
  length in Ref.~\onlinecite{hicks2018aflow}} as the result of finite
precision errors (the danger of these increases as the density of the
integration grid increases).

In this brief report, an algorithm for generating, and subsequently
symmetry-reducing, \kb-point grids is explained. This algorithm builds
on concepts such as Hermite Normal Form, Smith Normal Form, and the
connection between finite groups and integer matrices. These concepts
are briefly explained in the main text; for more details, see the
appendix and Ref.~\onlinecite{hart2008algorithm}.  The algorithm has
been implemented in an open-source code available at
\url{https://github.com/msg-byu/kgridGen} and
incorporated in version 6 of the VASP code.\cite{kresse1993ab} The algorithm has been incorporated into a code for generating generalized regular grids \url{https://github.com/msg-byu/GRkgridgen}.\cite{morgan2019generalized} 


\section{Generating grids}\label{sec:generateGrids}

\begin{figure}
  \includegraphics[width=1.0\linewidth]{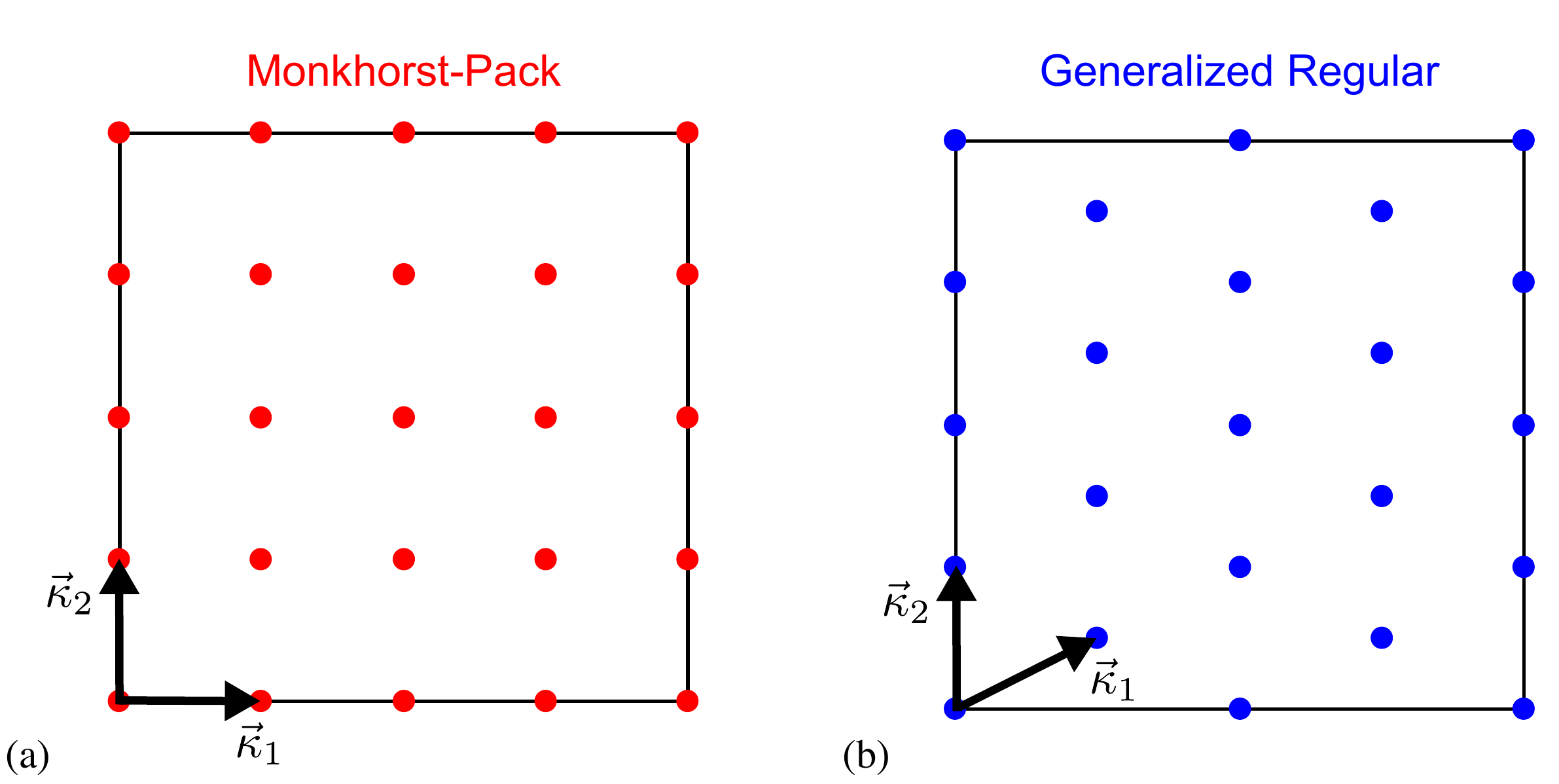}
  \caption{Two-dimensional example of a Monkhorst-Pack grid (a) and a generalized regular grid (b). The two grids have the same \kb-point density, but different grid-generating
  vectors $\vec{\kappa}_1$ and $\vec{\kappa}_2$. Both grids are commensurate with
  the reciprocal unit cell, shown as a black square. For Monkhorst-Pack grids, the
  matrix  $\mathbb{N}$ in Eq.~\ref{eqn:rkn} is integer and diagonal.
  In contrast, for generalized regular grids, $\mathbb{N}$ is not necessarily diagonal but is any invertible integer matrix.}
  \label{fig:mp-gr-grids}
\end{figure}

\begin{figure}
  \includegraphics[width=1.0\linewidth]{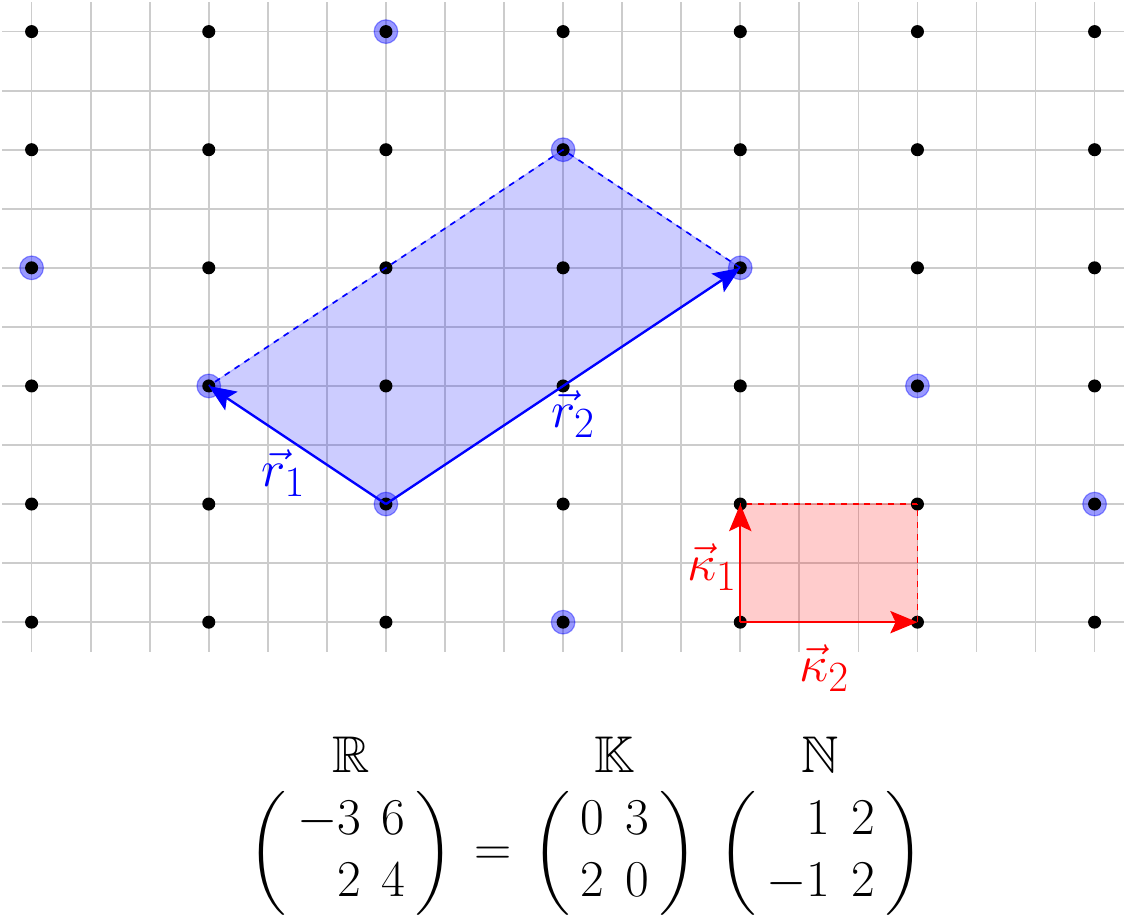}
  \caption{An example of the integer relationship between the reciprocal
    lattice vectors $\mathbb{R}$ and the grid generating vectors
    $\mathbb{K}$. In the picture, the grid generating vectors,
    $\vec{\kappa}_1$ and $\vec{\kappa}_2$, the columns of
    $\mathbb{K}$, define a lattice of points, four of which are inside
    the unit cell (blue parallelogram) of $\mathbb{R}$. Note that in
    the most general case, the relationship between the two lattices,
    $\mathbb{N}$ need not be diagonal (as it is for
    Monkhorst-Pack\cite{monkhorst1976special} \kb-point grids.)}
  \label{fig:rkn}
\end{figure}

As demonstrated in Fig.~\ref{fig:rkn}, every
uniform sampling of a reciprocal unit cell can be expressed through
the simple integer relationship
\begin{equation}
\mathbb{R}=\mathbb{KN}\label{eqn:rkn}
\end{equation}
where $\mathbb{R}$, $\mathbb{K}$, and $\mathbb{N}$ are $3\times 3$
matrices; the columns of $\mathbb{R}$ are the reciprocal lattice
vectors, and the columns of $\mathbb{K}$ are the \kb-point grid
generating vectors. Put simply, $\mathbb{N}$ describes the integer
linear combination of vectors of $\mathbb{K}$ that are equivalent to
$\mathbb{R}$. One obtains Monkhorst-Pack grids (regular grids) when
$\mathbb{N}$ is an integer, diagonal matrix. More generally, when
$\mathbb{N}$ is an invertible, integer matrix, one obtains
\emph{generalized regular} grids. Examples of Monkhorst-Pack and generalized
regular grids are given in Fig. \ref{fig:mp-gr-grids}. We use $R$ to refer to the infinite lattice of
points defined by integer linear combinations of $\mathbb{R}$,
and $K$ to refer to the lattice of points defined by $\mathbb{K}$.

With no loss of generality, a new basis for the lattice $K$ can be
chosen (a different, but equivalent, $\mathbb{K}$) so that
$\mathbb{N}$ is a lower triangular matrix in Hermite normal form
(HNF) \cite{storjohann1994computation}. (See Sec.~II-A of Ref.~\onlinecite{hart2008algorithm} for a
brief introduction to HNF.) HNF is a lower-triangular canonical matrix
form, where the entries below the diagonal are non-negative and strictly less
than the diagonal entry in the same row. Code for converting integer matrices to Hermite Normal Form
is available at \verb=https://github.com/msg-byu/symlib= in the
\verb=rational_mathemematics= module.

A \kb-point integration grid is the set of points of the lattice $K$ that lie
inside one unit cell (one fundamental domain) of the reciprocal
lattice $R$. We refer to this finite subset of $K$ as $K_\alpha$ (See
Fig.~\ref{fig:rkn}; black dots are $K$, dots inside the blue
parallelogram comprise $K_\alpha$.) The number of points that lie
within one unit cell of $R$ is given by $|\det(\mathbb{N})|=n$.

How then does one generate these $n$ points?  If $\mathbb{N}$ is in
HNF, then the diagonal elements of $\mathbb{N}$ are three integers,
$a$, $c$, and $f$, such that $a\cdot c\cdot f=n$. A set of $n$
translationally distinct \footnote{If two points are translationally
  distinct, their difference cannot be an integer linear combination
  of the \emph{reciprocal} cell vectors; that is,$\;\vec k_i-\vec
  k_j\neq n\vec r_1+m\vec r_2+\ell\vec r_3$, for all integer values
  $n,m,\ell$. (${\vec r_i}$ are the columns of $\mathbb{R}$.)}  points
of the lattice $K$ can be generated by taking integer linear
combinations of the columns of $\mathbb{K}$:

\begin{figure}
  \includegraphics[width=.85\linewidth]{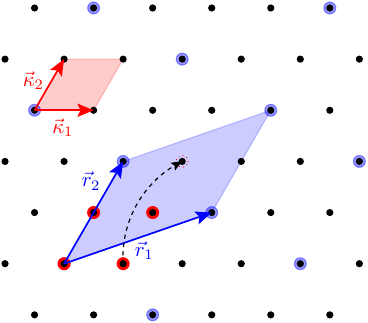}
\begin{center}
$\begin{array}{ccc}\\[0.05in]
\mathbb{K}=(\vec{\kappa}_1,\vec{\kappa}_2) = \left(\begin{array}{rr}
  1&\frac{1}{2}\\[4pt]0&\frac{\sqrt 3}{2}\end{array}\right)\\[0.3in]
  \end{array}$
$\begin{array}{ccc}
\mathbb{R}=\left(\begin{array}{rr} \frac{5}{2}&1\\[4pt]\frac{\sqrt
    3}{2}&\sqrt{3}\end{array}\right)&&\mathbb{N}=\left(\begin{array}{rr}
  2&0\\1&2\end{array}\right)

\end{array}$
\end{center}

\caption{An example of generating the points of $K$ (black lattice)
  that lie within one unit cell (blue parallelogram) of the lattice
  $R$ (blue lattice). The lattice $K$ is generated by the basis
  $\{\vec{\kappa}_1,\vec{\kappa}_2\}$ (columns of $\mathbb{K}$). The
  four points of $K_\alpha$ are generated by $\vec
  k=m_1\vec{\kappa}_1+m_2\vec{\kappa}_2$, where $0\leq m_1<2$,
  $\;0\leq m_2<2$. Note that the upper limits of $m_1$ and $m_2$ are
  the diagonals of $\mathbb{N}$ when it is expressed in HNF.}
  \label{fig:generateKgrid}
\end{figure}

\begin{equation}
\vec{k}=p\vec{\kappa}_1+q\vec{\kappa}_2+r\vec{\kappa}_3, \label{eqn:generateKgrid}
\end{equation}
where $p$, $q$, and $r$ are nonnegative integers such that
\begin{center}
$\begin{array}{c} 0\leq p<a\\[.05in] 0\leq q<c\\[.05in] 0\leq r<f.
\end{array}$
\end{center}
The $n$ points generated this way will not generally lie inside the
same unit cell, but they can be translated into the same cell by
expressing them in ``lattice coordinates'' (fractions of the columns
of $\mathbb{R}$, instead of Cartesian coordinates) and then reducing
the coordinates modulo 1 so that they all lie within the range $[0,
  1)$. This is illustrated by the dashed arrow in
  Fig.~\ref{fig:generateKgrid}.

Expressed as fractions of the lattice vectors of $R$, these four
points are:

\begin{center}
$\begin{array}{ccl} \vec{k}_1&=&(0,0)\\[.05in]
    \vec{k}_2&=&\left(0,\frac{1}{2}\right)\\[-.1in]
    \vec{k}_3&=&\left(\frac{1}{2},-\frac{1}{4}\right)\begin{array}[b]{c}
      \\[-.1in]\scriptstyle{\text{mod\ }1}\\[-.05in] \longrightarrow
\end{array} \left(\frac{1}{2},\frac{3}{4}\right)\\[.05in]
\vec{k}_4&=&\left(\frac{1}{2},\frac{1}{4}\right).
\end{array}$
\end{center}
Initially, $\vec k_3$ is not in the same unit cell as the other three
points; its first coordinate is not between 0 and 1. After reducing
the first coordinate modulo 1, $\vec k_3$ moves to an equivalent
position in the same unit cell as the other three points.

In summary, this first part of the algorithm generates $n$
translationally distinct points and translates them all into the first
unit cell of $R$. (It is not necessary to translate all the points
into the first unit cell, but it is convenient to do so as a first
step to translating them into the first Brillouin zone. The
translation into the first Brillouin zone is less trivial and is
discussed in Sec.~\ref{sec:BZ}.)

\begin{figure}
  \includegraphics[width=0.6\linewidth]{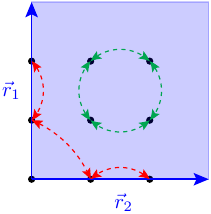}
\caption{An example of symmetry reducing a grid. The reciprocal unit
  cell is a square. This example assumes that the wavefunctions have
  square symmetry as well (the $D_4$ group, 8 operations). The example
  grid is a $3\times 3$ sampling of the reciprocal unit cell.  The
  point at $(0,0)$ is not equivalent to any of the other eight
  points. There are two sets of equivalent points, each set with 4
  points in the orbit, connected by red and green arrows,
  respectively. The points marked by red arrows are equivalent under
  horizontal, diagonal, and vertical reflections about the center of
  the square. The green-marked points are equivalent by 90$^\circ$
  rotations. Thus the nine points are reduced (or ``folded'') into
  \emph{three} symmetrically distinct points.
\label{fig:folding}}
\end{figure}

\section{Symmetry reduction of the grid}

In many cases the crystal will have some point group symmetries, and
these can be exploited to reduce the number of \kb-points where the
energy eigenvalues and corresponding wavefunctions need to be
evaluated.  The grid is reduced by applying the point group symmetries\footnote{In addition to the rotations, reflections,
and improper rotations of the crystal, inversion symmetry is also included by default. Even when the crystal itself
  does not have inversion symmetry, the electronic bands generally will. If, as in the case of magnetism, the inversion symmetry is broken, the inversion symmetry can be disabled in the code.} of
the crystal to
each point in the grid. For example, in Fig.~\ref{fig:folding}, the
points connected by green arrows will be mapped onto one another by
successive $90^\circ$ rotations. These four symmetrically equivalent
points lie on a 4-fold ``orbit'' (as do the points marked by the red
arrows). The point at the origin maps onto itself under all symmetry
operations and has an orbit of length 1.

For a grid containing $N_k$ points and a group (of rotation and
reflection symmetries) with $N_G$ operations, a naive algorithm for
identifying the symmetrically equivalent points and counting the
length of each orbit would be as follows: For each point
($\mathcal{O}(N_k)$ loop), compare all rotations of that point (a loop
of $\mathcal{O}(|G|)$) to all other points (another $\mathcal{O}(N_k)$
loop) to find a match; for a total computational complexity of
$\mathcal{O}(N^2_k N_G)$ (where $N_G$ is the number of rotation and
reflection symmetries). The algorithm is shown in pseudocode in Fig. \ref{fig:algorithm1}.
$N_G$ will never be larger than 48, but $N_k$ may be as large as
$50^3$ for extremely dense grids, so the $N_k^2$ complexity of this
naive approach is undesirable. But using group theory concepts (see
the Appendix for details), we can construct a hash table for the
points that reduces the complexity from $\mathcal{O}(N_k^2N_G$) to
$\mathcal{O}(N_k N_G)$ by eliminating the $k_j$ loop in Algorithm 1. The hash table makes a one-to-one association between the
ordinal counter (the index) of each point and its coordinates.

\begin{figure}[h]
{\algorithm
  \begin{tabbing}
    \textbf{Algorithm 1}\\[-.1in] \= mm \= mm \= mm \= mm \= mm \= mm
    \= mm \kill\\[.05in]
    \> uniqueCount $\longleftarrow 0$\\
    \>First[:] $\longleftarrow 0$\\
    \> Wt[:] $\longleftarrow 0$\\
    \>unique[:] $\longleftarrow 1$\\
    \>for each $\mathbf{k}_i\in K_\alpha$\\
    \>\> if unique[i] $\neq 1$ cycle \verb=#this=\\
    \>\> \verb=#point and all its symmetry equivalent=\\
    \>\> \verb=#points have already been indexed=\\
    \>\> uniqueCount++\\
    \>\> First[uniqueCount] $\longleftarrow$ i\\
    \>\> unique[i] $\longleftarrow 0$\\
    \>\> Wt[uniqueCount] $\longleftarrow 1$\\
    \>\> \verb=# Now mark all equivalent points=\\
    \>\>for each $\mathbf{k}_j\in K_\alpha$\\
    \>\>\> for each $g\in G$\\
    \>\>\>\> if $\mathbf{k}_j=g \cdot \mathbf{k}_i$\\
    \>\>\>\>\> Wt[uniqueCount]; ++\\
    \>\>\>\>\> unique[j] $\longleftarrow 0$
  \end{tabbing}
}
\cprotect\caption{The typical algorithm for reducing a grid by symmetry. In the algorithm,
\verb=uniqueCount= is a running counter of the number of unique
points and serves as the index of the orbit, and \verb=First= is a list of the indices of
the unique points, \verb=Wt= (weight) is the number of symmetrically equivalent versions of
each \kb-point in \verb=First= in the grid. \verb=unique= is an array of ones and zeros  where each element
corresponds to a \kb-point in $K_\alpha$. An element gets set to zero when the corresponding \kb-point is equivalent to another
 \kb-point, or when the \kb-point becomes the representative \kb-point of an orbit. This algorithm scales quadratically with the
 number of points in $K_\alpha$ and requires floating point comparisons between $\mathbf{k}_i$ and $\mathbf{k}_j$.}
\label{fig:algorithm1}
\end{figure}

The three coefficients $p,q,r$ in Eq.~(\ref{eqn:generateKgrid}) can be
conceptualized as the three ``digits'' of a 3-digit mixed-radix number
$pqr$ or as the three numerals shown on an odometer with three
wheels. The ranges of the values are $0\leq p<d_1$, $0\leq q<d_2$, and
$0\leq r<d_3$, where $d_1,d_2,d_3$ are the ``sizes'' of the wheels, or
in other words, the \emph{base} of each digit. Then the mixed-radix
number is converted to base 10 as
\begin{equation}
x=p\cdot d_2\cdot d_1+q\cdot d_1+r. \label{eqn:hash}
\end{equation}

The total number of possible readings of the odometer is $d_3\cdot
d_2\cdot d_1$. So it must be the case that the number of \kb-points in
the cell is $n=d_3\cdot d_2\cdot d_1$. Each reading on the odometer is
a distinct point of the $n$ points that are contained in the
reciprocal cell. Via Eq.~(\ref{eqn:hash}) it is simple to convert a
point given in ``lattice coordinates'' as $(p,q,r)$ to a base-10
number $x$. The concept of the hash table is to use this base-10
representation as the index in the hash table. Without the hash table,
comparing two points is an $\mathcal{O}(N_k)$ search because one point
must be compared to every other point in the list to check for
equivalency. But with the hash function, no search is necessary---one simply maps the point $(p,q,r)$ to the index $x$ of the equivalent \kb-point in the hash table.

It is not generally the case that the coefficients $p,q,r$ for every
interior point of the unit cell obey conditions:

\begin{equation}
0\leq p<d_1,\quad 0\leq q<d_2,\quad 0\leq r<d_3\label{eqn:interior}
\end{equation}
(Fig. \ref{fig:generateKgrid} shows an example where the interior
points do not meet these conditions.) These conditions hold only for a
certain choice of basis. That basis is found by transforming the
matrix $\N$ in Eq.~(\ref{eqn:rkn}) into its Smith Normal Form \cite{storjohann1994computation}, $
\mathbb{D} = \mathbb{A N B}$. By elementary row and column operations,
represented by unimodular matrices $\mathbb{A}$ and $\mathbb{B}$, it
is possible to transform $\N$ into a diagonal matrix $\D$, where each
diagonal element divides the ones below it: $d_{11}|d_{22}|d_{33}$,
and $d_{11}\cdot d_{22}\cdot d_{33}=n=\left|\N\right|$ (the notation
$i|j$ means that $i$ is divisible by $j$). As explained in the
appendix (Sec.~\ref{appendix:SNF}), when $\N$ is expressed in Smith
normal form (SNF) and the interior points of the reciprocal cell are
expressed as linear combinations of the grid generating vectors
$\mathbb{K}$, then the coordinates (coefficients) of the interior
points will obey Eq.~\ref{eqn:interior}. When these conditions are
met, the hashing algorithm discussed above (in particular,
Eq.~\ref{eqn:hash}) becomes possible. This enables the
$\mathcal{O}(N_k)$ algorithm, shown in Fig. \ref{fig:algorithm2}.
\begin{figure}
\centering
{\algorithm
  \begin{tabbing}
    \textbf{Algorithm 2}\\[-.1in] \= mm \= mm \= mm \= mm \= mm \= mm
    \= mm \= mm \kill\\[.05in]
    \> uniqueCount $\longleftarrow 0$\\ 
    \> hashTable[:] $\longleftarrow 0$\\
    \> First[:] $\longleftarrow 0$\\
    \> Wt[:] $\longleftarrow 0$\\
    \> for each $\mathbf{k}_i \in K_\alpha$\\
    \>\> indx $\longleftarrow \K^{-1}\A\D\cdot \mathbf{k}_i$\\
    \>\> if hashTable[indx] $\neq 0$ cycle \verb=#this=\\
    \>\> \verb=#point and all its symmetry equivalent=\\
    \>\> \verb=#points have already been indexed=\\
    \>\> uniqueCount++\\
    \>\> hashtable[indx] $\longleftarrow$ uniqueCount\\
    \>\> First[uniqueCount] $\longleftarrow$ i\\
    \>\> Wt[uniqueCount] $\longleftarrow$ 1\\
    \>\> \# \verb=Now mark all equivalent points=\\
    \>\> for each $g \in G$\\
    \>\>\> $\mathbf{k}_\textrm{rot} \longleftarrow g\cdot \mathbf{k}_i$\\
    \>\>\> indx $\longleftarrow \K^{-1}\A\D\cdot \mathbf{k}_\textrm{rot}$\\
    \>\>\> if hashtable[indx] == 0\\
    \>\>\>\> hashtable[indx] $\longleftarrow$ uniqueCount\\
    \>\>\>\> Wt[uniqueCount]++
  \end{tabbing}
}
\cprotect\caption{Our algorithm that reduces a grid to a set of symmetrically distinct \kb-points.
In the algorithm, \verb=uniqueCount= is a running counter of the number of unique
points and serves as the index of the orbit, \verb=First= is a list of the indices of the unique
points, \verb=Wt= (weight) is the number of symmetrically equivalent versions of
each \kb-point in \verb=First= in the grid, and \verb=hashTable= is a hash table
that points from the position of a \kb-point in $K_\alpha$ to
the index of its orbit. In contrast to Algorithm 1, this algorithm scales linearly with the
number of points in the grid $K_\alpha$ and does not
require floating point comparisons.}
\label{fig:algorithm2}
\end{figure}

\section{Moving points into the first Brillouin zone}
\label{sec:BZ}
For accurate DFT calculations, it is best if the energy eigenvalues
(electronic bands) are evaluated at \kb-points inside the first
Brillouin zone, so our algorithm includes a step that finds the
translationally equivalent grid points in the Brillouin zone. (In
principle, the electronic structure $E(\mathbf k)$ should be the same in every unit
cell, but numerically the periodicity of the electronic bands is only
approximate, becoming less accurate for \kb-points in unit cells
farther from the origin.)

The first Brillouin zone of the reciprocal lattice is simply the
Voronoi cell centered at the origin---all \kb-points in the first
Brillouin zone are closer to the origin than to any other lattice
point. Conceptually, an algorithm for translating a \kb-point of the
integration grid into the first zone merely requires one to look at
all translationally equivalent ``cousins'' of the \kb-point and select
the one closest to the origin. But the number of translationally
equivalent cousins is countably infinite, so in practice,
the set of cousins must be limited only to those near the origin.

\begin{figure}
\includegraphics[width=\linewidth]{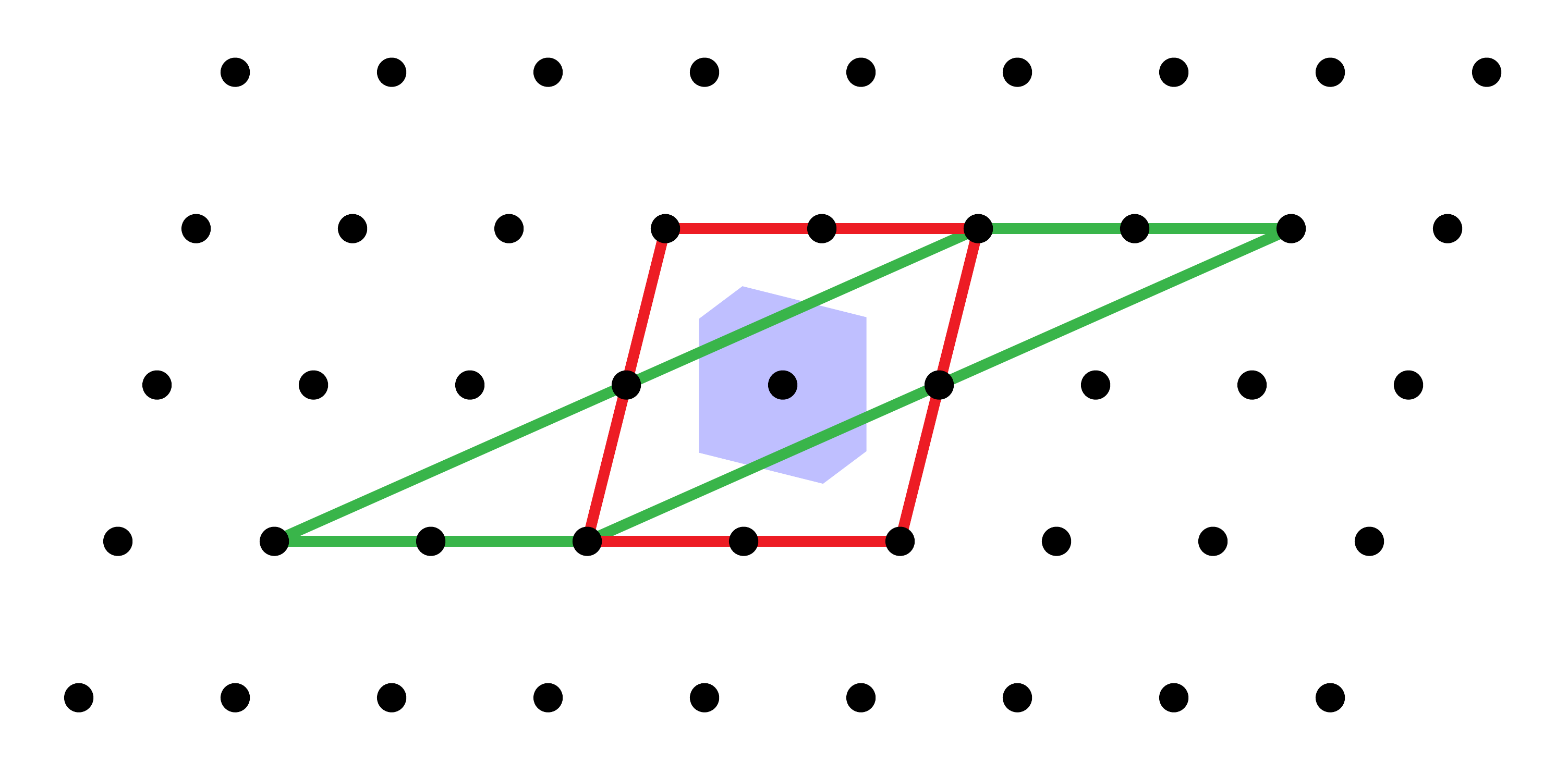}
\caption{A two-dimensional example of the ``closest cousin" guarantee. The Brillion zone (blue) will be completely contained in the
  union of 4 basis cells around the origin (shown in red) when the
  basis vectors are chosen to be as short as possible (the so-called
  Minkowski basis). On the other hand, if the basis is not Minkowski reduced, regions of
  the Brillouin zone may lie outside the union of the 4 basis cells (depicted by the cell in green). A proof is given in the appendix (Sec VII-A).}
\label{fig:8cells}
\end{figure}

How can we select a set of cousins near the origin that is guaranteed
to include the \emph{closest} cousin? The key idea is illustrated in
two-dimensions in Fig.~\ref{fig:8cells}. In three-dimensions, if the
basis vectors of the reciprocal unit cell are as short as possible
(the so-called Minkowski-reduced basis \cite{nguyen2009low}), then the eight unit cells
that all share a vertex at the origin \emph{must} contain the Brillouin zone. In other
words, the boundary of this union of eight cells is guaranteed to
circumscribe the first Brilloun zone (i.e., the Voronoi cell
containing the origin). A proof of this ``8 cells'' conjecture is
given in the Appendix (Sec VII-A). The steps for moving \kb-points into the
Brillouin zone are as follows:

    \begin{enumerate}
    \item Minkowski-reduce the reciprocal unit
      cell\cite{nguyen2009low} (i.e., find the basis with the shortest
      basis vectors\footnote{Our Fortran code for computing the
        Minkowski reduced basis is available at
        \texttt{https://github.com/msg-byu/symlib} in the
        \texttt{vector\_matrix\_utilities} module.})
    \item For each \kb-point in the reduced grid, find the
      translation-equivalent cousin in each of the eight unit cells
      that have a vertex at the origin.
      \item From these eight cousins, select the one closest to the
        origin.
      \end{enumerate}

\section{Conclusion}
We have developed an algorithm that i) generates generalized regular
\kb-point grids, ii) reduces the grids by symmetry, and iii) maps the points
of the reduced grids into the first Brillouin zone. Whereas the typical algorithm
for generating and reducing \kb-point grids scales quadratically with the number
of \kb-points, this algorithm scales linearly. The improved scaling becomes essential when one generates and symmetry reduces all combinatorically possible generalized regular grids
at a given \kb-point density, in order to select the one with the fewest number of
reduced \kb-points.\cite{morgan2019generalized}

The algorithm is also useful because it relies primarily
on integer-based operations, making it more robust than typical
floating point-based algorithms that are prone to finite precision
errors. Mapping the grid to the first Brillouin zone is more efficient due to a proof that
limits the search for translationally equivalent \kb-points to the eight unit cells having a
vertex at the origin. The algorithm has been incorporated into version 6 of
VASP.\cite{kresse1993ab}

\section{Acknowledgments} We are grateful to Martijn Marsmann who found a bug
in an early version of the code that spoiled the ${\cal O}(N_k)$
scaling and who implemented the algorithm in VASP, version 6. Tim
Mueller also contributed by pointing out a simpler approach to
generating all the \kb-points interior to the unit cell of $R$ than we
originally proposed. GLWH, WSM, and JJJ are grateful for financial
support from the Office of Naval Research (MURI N00014-13-1-0635).

\section{Appendix}\label{appendix:SNF}
\subsection{Proof limiting Brillouin zone location}

Given a point $x$ in the space, we will use the term \textit{cousin}
for a point $x'$ which differs from $x$ by an element of the
lattice---i.e., a coset representative or a lattice-translation
equivalent point.

Let $R$ be a basis. Let $U_R$ denote the union of $2^d$ basis cells
around the origin---the set of points which are expressible in terms
of the basis $R$ with all coefficients having absolute value $<$
1. Let $V$ denote the Voronoi cell (Brillouin zone)---the set of all
points in the space which are closer to the origin than any other
lattice point. Note that $U_R$ depends on the basis $R$, but $V$
depends only on the lattice. Note also that both $U_R$ and $V$ are
convex sets.

We claim (in two and three dimensions) that if $R$ is a Minkowski
basis, then $V \subseteq U_R$. We shall argue by contrapositive--- if
$V \nsubseteq U_R$, then the basis is not Minkowski reduced.

If $V \nsubseteq U_R$ then $V$ must intersect the boundary of $U_R$,
so there exist points on the boundary of $U_R$ which are closer to the
origin than to any other lattice points. Equivalently, those points
are closer to the origin than any of their cousins.

Note that among the cousins of any point on the boundary of $U_R$,
there is always a closest to the origin. But usually points on the
boundary will have closer cousins in the interior. But if $V
\nsubseteq U_R$ there must be points on the boundary which have no
closer cousins in the interior of $U_R$. In other words, there are
points (at least one) on the boundary such that \textit{all} of its
cousins in the interior of $U_R$ are farther from the origin.

\subsubsection{2D argument}

Let $\vec{r}_1$ and $\vec{r}_2$ be basis elements of $R$. Assuming
that $V \nsubseteq U_R$ there must be a point $x$ on the boundary of
$U_R$ whose cousins are all farther from the origin than $x$.

\begin{figure}
\includegraphics[width=\linewidth]{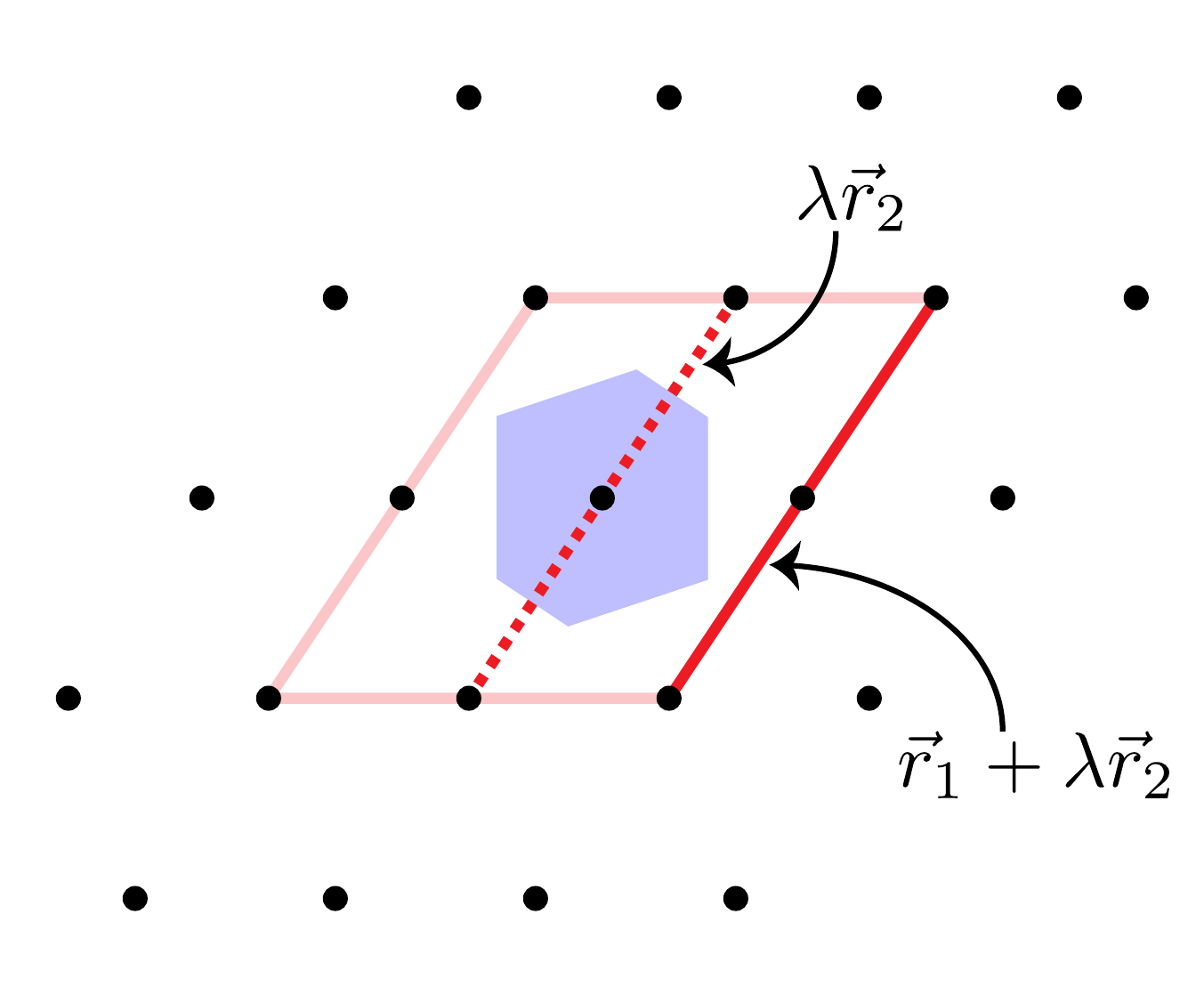}
\caption{Each point along the boundary of $U_R$ has at least one
  interior cousin closer to the origin when $R$ is Minkowski
  reduced. For the points along the edge in red, these interior
  cousins are the points along the dashed red line.}
\label{fig:bz-boundary}
\end{figure}

Without loss of generality (re-label the basis if necessary), we may
express one of the bounding edges of $U_R$ as $x = \vec{r}_1 + \lambda
\vec{r}_2$ where $\lambda \in [-1,1]$. One of its interior cousins is
$x' = \lambda \vec{r}_2$, which is illustrated in
Fig. \ref{fig:bz-boundary}. We have (since $x'$ must be farther from
the origin)
\begin{align*}
x^2 &< x'^2 \\ (\vec{r}_1 + \lambda \vec{r}_2)^2 &< (\lambda
\vec{r}_2)^2 \\ \vec{r}_1^2 + 2 \lambda \vec{r}_1 \cdot \vec{r}_2 +
\lambda^2 \vec{r}_2^2 &< \lambda^2 \vec{r}_2^2 \\ \vec{r}_1^2 &< -2
\lambda \vec{r}_1 \cdot \vec{r}_2
\end{align*}
Since the expression on the left-hand side is greater than zero, the
expression on the right-hand side must be also and taking the absolute
value of both sides does not change the inequality:
\begin{equation*}
|\vec{r}_1^2| < |-2 \lambda \vec{r}_1 \cdot \vec{r}_2| \implies
|\vec{r}_1|^2 < 2 |\lambda| | \vec{r}_1 \cdot \vec{r}_2|.
\end{equation*}
Considering the worst case scenario of $\lambda = 1$ gives
\begin{equation}
\frac{|\vec{r}_1|}{2} < \frac{| \vec{r}_1 \cdot
  \vec{r}_2|}{|\vec{r}_1|},
\end{equation}
which violates the condition of a Minkowski basis $|\vec{r}_1 \cdot
\vec{r}_2|/|\vec{r}_1| < |\vec{r}_1|/2$. The remaining three
boundaries are similar to the one just considered, the only
differences being permutations of the basis elements $\vec{r}_1$ and
$\vec{r}_2$ and possibly changes of sign. When applying the same
reasoning to the other edges we arrive at the same
contradiction. Hence, the points on the boundary of $U_R$ are closer
to the origin than interior cousins, $V \nsubseteq U_R$, only when the
basis $R$ is not Minkowski reduced. If $R$ is Minkowski reduced, all
points on the bounday of $U_R$ have interior cousins that lie closer
to the origin and $V \subseteq U_R$.

\subsubsection{3D argument}

Let $\vec{r}_1$, $\vec{r}_2$, and $\vec{r}_3$ be the basis elements of
$R$, and suppose (relabeling the basis vectors if necessary) that $x =
\vec{r}_1 + \lambda \vec{r}_2 + \delta \vec{r}_3$ (where $\lambda$ and
$\delta$ are elements of $[-1,1]$) is a point on the boundary of $U_R$
which is closer to the origin than are its interior cousins.

\begin{figure}
\includegraphics[width=.8\linewidth]{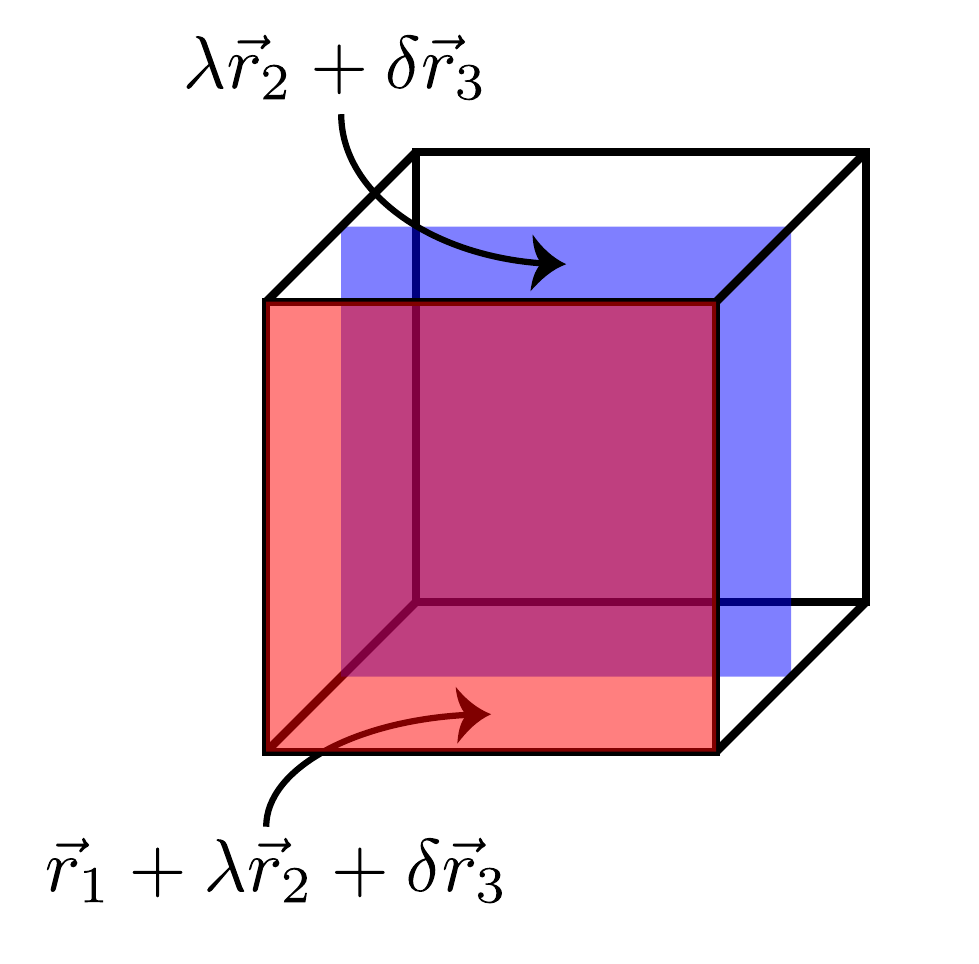}
\caption{Each point along the boundary of $U_R$, the edges of which
  are shown in black, has at least one interior cousin closer to the
  origin when $R$ is Minkowski reduced. For the points on the bounding
  plane in red, the interior cousins are the points on the plane in
  blue. (The origin is contained in the blue plane.)}
\label{fig:bz-boundary-3d}
\end{figure}

One of those cousins is a plane through the origin $x' = \lambda
\vec{r}_2 + \delta \vec{r}_3$. The boundary and cousin planes are
shown in Fig. \ref{fig:bz-boundary-3d}. Thus
\begin{align*}
x^2 &< x'^2 \\ (\vec{r}_1 + \lambda \vec{r}_2 + \delta \vec{r}_3)^2 &<
(\lambda \vec{r}_2 + \delta \vec{r}_3)^2.
\end{align*}
Simplifying this expression gives
\begin{equation}
\vec{r}_1^2 < -2 \lambda \vec{r}_1 \cdot \vec{r}_2 - 2 \delta
\vec{r}_1 \cdot
\end{equation}
Since the expression on the left-hand side is greater than zero, the
expression on the right-hand side must be also and taking the absolute
value of both sides does not change the inequality:
\begin{equation} \label{eq:bz-triag}
|\vec{r}_1^2| < |-2 \lambda \vec{r}_1 \cdot \vec{r}_2 - 2 \delta
\vec{r}_1 \cdot \vec{r}_3| \implies |\vec{r}_1|^2 < |2 \lambda
\vec{r}_1 \cdot \vec{r}_2 + 2 \delta \vec{r}_1 \cdot \vec{r}_3 |
\end{equation}

To simplify the right-hand side of Eq.~\ref{eq:bz-triag} the triangle
inequality is used, making it more likely that the inequality is
satisfied:
\begin{align} \label{eqn:proof1}
 |\vec{r}_1|^2 &< |2 \lambda \vec{r}_1 \cdot \vec{r}_2 + 2 \delta
|\vec{r}_1 \cdot \vec{r}_3 | \nonumber \\ \vec{r}_1|^2 &< 2 |\lambda|
||\vec{r}_1 \cdot \vec{r}_2 | + 2 |\delta| |\vec{r}_1 \cdot \vec{r}_3|
\end{align}

Since the expression in Eqn. \ref{eqn:proof1} does not depend on the
sign of $\lambda$ or $\delta$, we can restrict $\lambda$ and $\delta$
to positive values within $[0,1]$ without loss of generality. Consider
now \textit{another} cousin that lies within $U_R$ and on the same
plane as $x'$: $x'' = (\lambda - 1)\vec{r}_2 + (\delta -
1)\vec{r}_3$. Repeating the same process for $x'$ with $x''$ gives
\begin{equation} \label{eqn:proof2}
|\vec{r}_1|^2 < 2 |\lambda - 1| |\vec{r}_1 \cdot \vec{r}_2 | + 2
|\delta - 1| |\vec{r}_1 \cdot \vec{r}_3|
\end{equation}
Combining Eqns. \ref{eqn:proof1} and \ref{eqn:proof2} gives
\begin{align} \label{eqn:proof3}
|\vec{r}_1|^2 &< (|\lambda| + |\lambda - 1|) |\vec{r}_1 \cdot
|\vec{r}_2| + (|\delta| + |\delta - 1|) |\vec{r}_1 \cdot \vec{r}_3|
|\nonumber \\ \vec{r}_1| &< \frac{|\vec{r}_1 \cdot
|\vec{r}_2|}{|\vec{r}_1|} + \frac{|\vec{r}_1 \cdot
|\vec{r}_3|}{|\vec{r}_1|}
\end{align}
Assuming $\{\vec{r}_1, \vec{r}_2, \vec{r}_3\}$ forms a Minkowski
basis, and plugging in the largest possible values for all quantities
under this assumption on the right-hand side of Eqn. \ref{eqn:proof3}
gives the contradiction $|\vec{r}_1| < |\vec{r}_1|$. The remaining
seven bounding planes are similar to the one just considered, the only
differences being permutations of the basis elements $\vec{r}_1$,
$\vec{r}_2$, and $\vec{r}_3$ and changes of sign. We arrive at the
same contradiction when applying the same reasoning to the other
planes. Hence, the points on the boundary of $U_R$ are closer to the
origin than interior cousins, $V \nsubseteq U_R$, only when the basis
$R$ is not Minkowski reduced. If $R$ is Minkowski reduced, all points
on the boundary of $U_R$ have interior cousins that lie closer to the
origin and $V \subseteq U_R$.

\subsection{Groups, Matrices, and Lattices in Smith Normal Form.}
The discussion below is limited to three-dimensions though the
arguments easily generalize to higher dimensions. The purpose of the
discussion below is to help the reader make the connection between
groups and integer matrices. The Smith Normal Form is a key concept to
make this connection.

 In this discussion, we show that we can associate a single, finite
 group with the lattice sites within one tile (i.e., one unit cell) of
 a superlattice. In our application, this tile is the unit cell of the
 grid generating vectors and the superlattice is the reciprocal
 cell. The association between the group and the lattice sites is a
 homomorphism that maps each lattice site to an element of the
 group. If two points are translationally equilavent (same site but in
 two different tiles) they will map to the same element of the
 group. This homomorphism is the key ingredient to constructing the
 hash function (see Eq.~\ref{eqn:hash}) that enables a perfect hash
 table where points are listed consecutively, from 1 to $N$. In what
 follows, we explain in detal how this association is made, i.e., we
 detail how one finds this homomorphism.

\subsubsection{Groups in Smith Normal Form}
\label{appendix:groups}
Begin with the simplest case.  Let $\mathbb{N}$ be a non-singular $3
\times 3$ matrix of integers.  Its columns represent the basis for a
subgroup ${\cal L}_N$ of the group $\mathbf{Z}^3$ (where $\mathbf{Z}$
is the set of all integers, and the group operation is addition).  The
two latttices whose symmetries are represented by these two groups are
the ``simple cubic'' lattice of all points with all integer
coordinates and its \emph{superlattice}
\footnote{In the mathematical literature, and in some of the
  crystallography literature, these ``superlattices'' are referred to
  as sublattices. The group associated with a ``superlattice'' is a
  \emph{subgroup} of the group associated with the parent lattice.
  Although this nomenclature (subgroups, sublattices) is more correct
  from a mathematical or group theory point of view, we follow the
  nomenclature typically seen in the physics literature where a
  lattice or a structure whose volume is larger than that of the
  parent is referred as a superlattice.}  whose basis is given by the
columns of $\mathbb{N}$.  Since $\mathbf{Z}^3$ and its subgroups are
Abelian, we know that all the subgroups are \emph{normal} so there
exists a quotient group $G=\mathbf{Z}^3/{\cal L}_N$, and that group is
\emph{finite}.

Note that the \emph{cosets} which form the elements of that quotient
group are simply the distinct translates of the lattice ${\cal L}_N$
within $\mathbf{Z}^3$.  In fact, each coset has exactly one
representative in each unit cell, so the order of $G$ is equal to the
volume of a unit cell (the absolute value of the determinant of
$\mathbb{N}$).  Since the quotient group $G$ is finite, and Abelian,
it must be a direct sum of cyclic groups (by the Fundamental Theorem
of Finite Abelian Groups).

One canonical form for direct sums of groups is called Smith Normal
Form, where the direct summands are ordered so that each summand
divides the next.  In other words, $G\cong \mathbf{ Z}_{m_1}\oplus
\mathbf{Z}_{m_2}\oplus \cdots \oplus \mathbf{ Z}_{m_k}$ where $m_1 |
m_2 | \dots m_{k-1} | m_k$ and (of course) $\prod m_i=|G|$.  Any
finite Abelian group can be uniquely written in this form.
(Isomorphic groups will yield the same ``invariant factors'' $m_1$,
$m_2$,\dots,$m_k$ when written in this form.)

Note that, since $G=\mathbf{Z}^3/{\cal L}_N$, there must be a
homomorphism from $\mathbf{Z}^3$ onto $G$, having ${\cal L}_N$ as its
kernel. In other words, ${\cal L}_N=\{p\in \mathbf{Z}^3 \>:\>
\psi(p)=0\}$.  Our task is to find the direct-sum representation of
the quotient group $\mathbf{Z}^3/{\cal L}_N$, and also to find the
homomorphism $\psi$ which maps the points of $\mathbf{Z}^3$ onto the
group (in such a way that $\psi(p)=0$ iff $p\in {\cal L}_N$). This
allows us to work with the elements of the group as proxies for the
\kb-points inside the reciprocal cell.

\subsubsection {Matrices in Smith Normal Form}
%
There is a useful connection between the SNF for Abelian groups and
the SNF for integer matrices.  As the reader may infer, the SNF form
of the basis matrix $\mathbb{N}$ effectively tells one how to
represent the quotient group $\mathbf{Z}^3/{\cal L}_N$ as a direct sum
of cyclic groups in Smith Normal Form, and, as shown in the following
section, the row operations used to create the SNF of $\mathbb{N}$
give the homomorphism $\psi$ suggested above.

\subsubsection{The connection between SNF groups and SNF matrices}
In the matrix case, since the operations are elementary row and column
operations, we have $\mathbb{D}=\mathbb{ANB}$ where $\mathbb{A}$ and
$\mathbb{B}$ are integer matrices with determinant $\pm 1$
representing the accumulated row operations and column operations
respectively.  The matrix $\mathbb{D}$ is completely determined by
$\mathbb{N}$, but the matrices $\mathbb{A}$ and $\mathbb{B}$ depend on
the algorithm used to arrive at the Smith Normal Form of $\mathbb{N}$.
A different implementation might yield $\mathbb{D}=\mathbb{A'NB'}$
(same $\mathbb{N}$ and same $\mathbb{D}$, but different $\mathbb{A}$
and $\mathbb{B}$).

Note that, since $\mathbb{B}$ represents elementary column operations,
the product $\mathbb{NB}$ simply represents a change of basis from
$\mathbb{N}$ to a new basis $\mathbb{N}'=\mathbb{NB}$.  In other
words, the columns of $\mathbb{N}'$ are still a basis for ${\cal
  L}_N$.  But the new basis has the property that
$\mathbb{AN}'=\mathbb{D}$. That means that every element $\vec
w=\mathbb{N}'\vec z$ of ${\cal L}_N$ (where $\vec z$ is some element
of $\mathbb{Z}^3$) will satisfy the equation $\mathbb{A}\vec w =
\mathbb{D}\vec z$ $= \left(\begin{matrix}\mathbb D_{11} z_1\\ \mathbb
  D_{22}z_2\\ \mathbb D_{33}z_3\end{matrix}\right)$.  In other words,
  $\mathbb{A}\vec w$ will be a vector whose entries are
  \emph{multiples} of the corresponding diagonal entries in
  $\mathbb{D}$.

To put it another way, define $*$ to be the operation that maps $\vec
x=\left(\begin{matrix}x_1\\ x_2\\ x_3\end{matrix}\right)$ in $\Z^3$ to
  $\vec x^*= \left(\begin{matrix}x_1\;{\rm
      (mod\>\mathbb{D}_{11})}\\ x_2\;{\rm
      (mod\>\mathbb{D}_{22})}\\ x_3\;{\rm
      (mod\>\mathbb{D}_{33})}\end{matrix}\right)^T$.

  Then we have shown $\vec w\in {\cal L}_N$ iff $(\A w)^*=(0,0,0)$
  (the zero-element in the group $G_0={\mathbb
    Z}_{\mathbb{D}_{11}}\oplus {\mathbb Z}_{\mathbb{D}_{22}} \oplus
  {\mathbb Z}_{\mathbb{D}_{33}}$.

That suggests we let $\psi(\vec w)=(\mathbb{A}\vec w)^*$, a
homomorphism from $\mathbf{Z}^3$ onto the direct-sum $G_0$.  Then,
since that homomorphism is easily shown to be onto, and its kernel is
${\cal L}_N$, we see (by the First Isomorphism Theorem of group
theory) that $G_0\cong \mathbf{Z}^3/{\cal L}_N$, and $\psi$ is
precisely the homomorphism we sought.

Thus we have connected the two versions of SNF.  The matrix algorithm
provides the SNF description of the quotient group by the diagonal
entries in $\mathbb{D}$, and the transition matrix $\mathbb{A}$
provides the homomorphism which maps the parent lattice onto the
group.

\bigskip

\noindent {\bf An example.}  Let $\mathbb{N}=\left(\begin{matrix}
  1&2&-1\\1&4&-3\\0&2&4\end{matrix}\right)$.  This describes a lattice
  ${\cal L}_N$ which contains the points $\vec
  p_1=\left(\begin{matrix}1\\ 1\\ 0\end{matrix}\right)$, $\vec
    p_2=\left(\begin{matrix}2\\ 4\\ 2\end{matrix}\right)$, and $\vec
      p_3=\left(\begin{matrix}-1\\ -3\\ 4\end{matrix}\right)$, and
        \emph{all} the points which are integer linear combinations of
        those three points.  The matrix $\mathbb{N}$ has determinant
        12, which must be the volume of each lattice tile---and it is
        also the order of the quotient group $\mathbf{Z}^3/ {\cal
          L}_N$.

Using the SNF algorithm to diagonalize this basis matrix, we find
$\mathbb{D}=\mathbb{ANB}$ where
$\mathbb{D}=\left(\begin{matrix}1&0&0\\ 0&2&0\\ 0&0&6\end{matrix}\right)$,
  with
  $\mathbb{A}=\left(\begin{matrix}0&1&0\\ 0&0&1\\ 1&-1&-2\end{matrix}\right)$
    and
    $\mathbb{B}=\left(\begin{matrix}1&7&11\\ 0&-1&-2\\ 0&1&1\end{matrix}\right)$.

Thus we now know that the quotient group is $G=\mathbf{Z}^3/ {\cal
  L}_N\cong \mathbf{Z}_1 \oplus \mathbf{Z}_2 \oplus \mathbf{Z}_6 \cong
\mathbf{Z}_2 \oplus \mathbf{Z}_6$.

\medskip
Further, from the matrix $\mathbb{A}$, we may obtain the homomorphism
projecting $\mathbf{Z}^3$ onto the quotient group, with kernel ${\cal
  L}_N$.  If $\vec w=\left(\begin{matrix}x\\ y\\ z\end{matrix}\right)$
  then $\mathbb{A}\vec
  w=\left(\begin{matrix}y\\ z\\ x-y-2z\end{matrix}\right)$ and thus

\begin{align*}
\psi(\vec w) &=(\mathbb{A}\vec w)^*\\ &= \left(\begin{matrix}y\>{\rm
    (mod\>}1)\\ z\>{\rm (mod\>}2)\\ x-y-2z\>{\rm
    (mod\>}6)\end{matrix}\right)^{T} \\ &= \bigl( z\> {\rm (mod \>}
2), x+5y+4z\> {\rm (mod \>} 6)\bigr)
\end{align*} (noting that anything
    mod 1 is zero).

\bigskip

Note that this homomorphism provides a different, but convenient, way
to describe the superlattice.  Since ${\cal L}_N$ is the kernel of
$\psi$, it is comprised of the points $(x,y,z)\in \mathbf{Z}^3$ which
satisfy the simultaneous congruences $z\equiv 0$ (mod 2) and
$x+5y+4z\equiv 0$ (mod 6).  We note that all three basis points $p_1$,
$p_2$ and $p_3$ satisfy these congruences, and thus so will all their
integer linear combinations (all points in ${\cal L}_N$).

\bigskip

\noindent {\bf Algorithmic variation.}  In the example we computed
above, a different application of the SNF matrix algorithm, with the
same $\N$, might have yielded the same diagonal matrix
$\mathbb{D}=\left(\begin{matrix}1&0&0\\ 0&2&0\\ 0&0&6\end{matrix}\right)$,
  but different
  $\mathbb{A}=\left(\begin{matrix}1&0&0\\ -5&3&1\\ -2&2&1\end{matrix}\right)$
    and
    $\mathbb{B}=\left(\begin{matrix}0&-1&2\\ 0&0&1\\ -1&-1&4\end{matrix}\right)$,
      which would change the homomorphism to $(x,y,z)\mapsto \bigl(
      -5x+3y+z\;({\rm mod}\;2),\; -2x+2y+z \;({\rm mod}\;6)\bigr) =
      \bigl( x+y+z\;({\rm mod}\;2),\; 4x+2y+z \;({\rm
        mod}\;6)\bigr).$\medskip The new homomorphism is different,
      since $(1,0,1)\mapsto (0,5)$ now, where previously
      $(1,2,3)\mapsto(1,5)$ (for example), but the \emph{kernel} is
      the same.  In fact the two homomorphisms are related via an
      automorphism of the group $G$.
\bigskip

%
\subsubsection{Non-integer lattices}

Now, what about the more complicated situation, where $\mathbb{N}$
represents a (possibly HNF) matrix describing the change from some
lattice other than the simple integer lattice $\mathbf{Z}^3$ to one of
its subgroups (superlattice)?

Then we have a basis $\mathbb{V}$ and lattice ${\cal L}_V$ and a basis
$\mathbb{W} = \mathbb{VN}$ for a (super) lattice ${\cal L}_W$.  Again,
the quotient group $G={\cal L}_V/{\cal L}_W$ is Abelian of order
$|$det$(\mathbb{N})|$.  Again, $G$ is a direct sum of cyclic groups
corresponding to the diagonal entries of $\mathbb{D} = \mathbb{A N B}$
(where $\mathbb{D}$ is the SNF of $\mathbb{N}$).

The only difference here is that the homomorphism $\psi$ provided by
$\mathbb{A}$ must depend on the basis $\mathbb{V}$ (which might even
be irrational).  Every point in ${\cal L}_V$ has the form $\vec x =
\mathbb{V}\vec w$ where $\vec w$ is a column of integers.  Then
$\psi(\vec x)=\mathbb{A}\vec w$ (modded by the corresponding entries
from $\mathbb{D} = \mathbb{A N B}$).  We could write it as $\psi(\vec
x)=\bigl(\mathbb{A} \mathbb{V}^{-1} \vec x \bigr)^*$ (with the entries
appropriately modularly reduced and transposed to a horizontal
vector).

\subsubsection{ Example: general (non-integer) lattices}

Suppose ${\cal L}_V$ is the lattice defined by (columns of) the basis
matrix $\mathbb{V}=\left(\begin{matrix}1&1/2&0\\ 0&\sqrt{3}/2
  &0\\ 0&0&2\end{matrix}\right)$, and ${\cal L}_W$ is the subgroup
  lattice defined by the basis marix $\mathbb{W} = \mathbb{V N}$ where
  $\mathbb{N}=\left(\begin{matrix}
    4&2&2\\2&2&2\\4&0&4\end{matrix}\right)$. In other words, one basis
    for ${\cal L}_W$ is given by the columns of
    $\mathbb{W}=\left(\begin{matrix}5&3&3\\ \sqrt{3}&\sqrt{3}&\sqrt{3}\\ 8&0&8\end{matrix}\right)$.

Reducing $\N$ to SNF yields
$$\mathbb{D}=\left(\begin{matrix}2&0&0\\ 0&2&0\\ 0&0&4\end{matrix}\right)=\left(\begin{matrix}1&0&-1\\ 1&-1&-1\\ -6&4&5\end{matrix}\right)\mathbb{N}
    \left(\begin{matrix}-2&-3&-2\\ 2&1&1\\ 1&1&1\end{matrix}\right).$$

Thus our quotient group is $G={\cal L}_V/{\cal L}_W\cong
\mathbf{Z}_2\oplus\Z_2\oplus\Z_4$ and
$\mathbb{A}=\left(\begin{matrix}1&0&-1\\ 1&-1&-1\\ -6&4&5\end{matrix}\right)$
  so $$\mathbb{A} \mathbb{V}^{-1}\;=\;
  \left(\begin{matrix}1&-\sqrt{3}/3&-1/2\\ 1&-\sqrt{3}&-1/2\\ -6&14\sqrt{3}/3
    &5/2\end{matrix}\right),$$ which provides our homomorphism
    $\psi(\vec x)=\bigl(\mathbb{A} \mathbb{V}^{-1}x\bigr)^*$ from
    ${\cal L}_V$ onto $G$.

If we let $\vec x=\left(\begin{matrix}2\\ \sqrt{3}
  \\ 2\end{matrix}\right)$ which is an element of ${\cal L}_V$ but not
  of ${\cal L}_W$, then $\mathbb{A}
  \mathbb{V}^{-1}x=\left(\begin{matrix} 0\\ -2
    \\ 7\end{matrix}\right)$ and $\psi(\vec x)=(0,0,3)\in G$ (after
    reducing the elements modulo 2, 2 and 4 respectively).  On the
    other hand, if we let $\vec x=\left(\begin{matrix}7\\ \sqrt{3}
      \\ 8\end{matrix}\right)$, which \emph{is} an element of ${\cal
        L}_W$ (the kernel), then $\mathbb{A} \mathbb{V}^{-1} \vec
      x=\left(\begin{matrix} 2\\ 0 \\ -8\end{matrix}\right)$ and so
        $\psi(\vec x)=(0,0,0)$, and $\vec x$ is in the group.

By this function $\psi$, the elements of ${\cal L}_V$ are all mapped
to elements of the group $G$ and, in particular, the elements of
${\cal L}_W$ are mapped to the zero element of the group.  Stated in
terms of the cosets, the entire set ${\cal L}_W$ is mapped to the zero
element of the group $G$, and each of the distinct translates of
${\cal L}_W$ (within ${\cal L}_V$) gets mapped to a different element
of the group.  We might think of this as decorating or \emph{labeling}
the elements of ${\cal L}_V$ in a periodic manner, using ${\cal L}_W$
to define the period, and using the elements of the group $G$ as the
labels.

\bibliography{master.bib}
\end{document}